\RequirePackage{amsmath}
\documentclass[runningheads]{llncs}
\usepackage[T1]{fontenc}
\usepackage{graphicx}

\usepackage[utf8]{inputenc}
\usepackage{stfloats}
\usepackage{url}
\usepackage{hyperref}
\usepackage{listings}
\usepackage{xspace}
\usepackage{subcaption}
\usepackage[ruled,vlined,linesnumbered]{algorithm2e}

\usepackage{array,booktabs}
\usepackage{tabu}
\usepackage{xparse} 
\usepackage{glossaries} 
\usepackage{subfiles} 
\usepackage{listings} 
\usepackage{newfloat} 
\usepackage{verbatim} 
\usepackage{fancyvrb} 
\usepackage[misc,geometry]{ifsym}

\usepackage[textsize=normalsize,textwidth=3cm]{todonotes}

\DeclareFloatingEnvironment[fileext=frm,placement={!ht},name=Listing]{listing}

\SetKwProg{Fn}{Plan}{}{}

\newglossaryentry{mas:jacamo}{name=JaCaMo,description={a platform for the development of multi-agent systems}}
\newglossaryentry{mas:jason}{name=Jason,description={AgentSpeak interpreter for programming agents}}
\newglossaryentry{mas:cartago}{name=CArtAgO,description={}}
\newglossaryentry{mas:moise}{name=\ensuremath{\mathcal{M}\textsc{oise}},description={}}
\newacronym{cnp}{CNP}{Contract Net Protocol}
\newacronym{mas}{MAS}{Multi-Agent System}
\newacronym{bdi}{BDI}{Belief Desire Intention}

\newacronym{lfc}{LFC}{Liverpool Formidable Constructors}
\newglossaryentry{fit}{name=FIT BUT,description={}}
\newglossaryentry{trg}{name=TRG,description={}}
\newglossaryentry{dtu}{name=GOAL-DTU,description={}}

\newacronym{mapc}{MAPC}{Multi-Agent Programming Contest}

\newglossaryentry{team:teamart}{name=\texttt{TeamArtifact},description={}}
\newglossaryentry{team:agentart}{name=\texttt{EISArtifact},description={}}
\newglossaryentry{team:translator}{name=\texttt{Translator},description={}}

\newacronym{fd}{FD}{Fast Downward}

\input{./snippet-code/jacamo.pygstyle}

\begin{document}

\title{LFC: Combining Autonomous Agents and Automated Planning in the Multi-Agent Programming Contest\thanks{Work supported by UK Research and Innovation, and EPSRC Hubs for Robotics and AI in Hazardous Environments: EP/R026092 (FAIR-SPACE), EP/R026173 (ORCA), and EP/R026084 (RAIN).}}

\titlerunning{LFC: Combining Autonomous Agents and Automated Planning}

\author{Rafael C. Cardoso$^{\textrm{(\Letter)}}$\orcidID{0000-0001-6666-6954}  \and
Angelo Ferrando\orcidID{0000-0002-8711-4670} \and
Fabio Papacchini} 

\authorrunning{R. C. Cardoso et al.}
%

\institute{University of Liverpool, Liverpool L69 3BX, United Kingdom\\
\email{\{rafael.cardoso,angelo.ferrando,fabio.papacchini\}@liverpool.ac.uk}}

\maketitle

\begin{abstract}
The 2019 Multi-Agent Programming Contest introduced a new scenario, Agents Assemble, where two teams of agents move around a 2D grid and compete to assemble complex block structures. In this paper, we describe the strategies used by our team that led us to achieve first place in the contest. Our strategies tackle some of the major challenges in the 2019 contest: how to explore and build a map when agents only have access to local vision and no global coordinates; how to move around the map efficiently even though there are dynamic events that can change the cells in the grid; and how to assemble and submit complex block structures given that the opposing team may try to sabotage us. To implement our strategies, we use the multi-agent systems development platform JaCaMo to program our agents and the Fast Downward planner to plan the movement of the agent in the grid. We also provide a brief discussion of our matches in the contest and give our analysis of how our team performed in each match.
    
\end{abstract}

\section{Introduction} 
\label{sec:intro}

The \acrfull{mapc} is an annual event to promote the use and improvement of agent programming languages. A number of teams face off in a challenging scenario that is made to encourage the use of agent techniques. The matches between teams (the clients) are played in a simulated environment (the server) where agents receive perceptions about the environment and can send actions to the server. The simulation occurs synchronously through simulation steps, that is, each agent is required to send its actions to the server before the step deadline (usually four seconds). The team with the highest score at the end of the match wins the round and is awarded some points. The winner of the contest is the team with the most points.

The 2019 \acrshort{mapc} introduced the Agents Assemble scenario. In this scenario, two teams of ten agents each compete to assemble complex block structures. Blocks can have different types and can be generated upon using the \emph{request} action at a block \emph{dispenser} of the desired type. The environment is represented in a 2D grid where each cell can contain different types of entities and/or terrain: a block; a block dispenser; an obstacle (prohibited cell); an agent; a goal cell; and/or an empty cell. Each agent only has local coordinates of the map (i.e., their position is always 0,0). Each match has three rounds, and the map used in each round is generated randomly. A complete description of the scenario is available in~\cite{MAPC2019}. For the remainder of this paper, we only describe the features of the scenario that are relevant to the strategy being discussed.

Our team, the \acrfull{lfc}, achieved first place in the 2019 \acrshort{mapc}. In this paper we discuss the main strategies that our agents used throughout the contest: grid exploration, identification of teammates, evaluation of good goal positions, creating and maintaining a global map, planning an efficient path, and task achievement. A key aspect that differentiates our team from others, even teams from past \acrshort{mapc}s, is the use of a classical planner to plan the movement of our agents in the grid. The \acrshort{lfc} source code, as used in the 2019 \acrshort{mapc}, is available at: <\url{https://github.com/autonomy-and-verification-uol/mapc2019-liv}>.

This paper is organised as follows. In the next section we describe the software architecture used in the development of our team, namely the \acrfull{mas} development platform JaCaMo~\cite{Boissier11} and the Fast Downward planner~\cite{Helmert06}. Section~\ref{sec:strategy} contains the strategies that we used to solve some of the major challenges present in the Agents Assemble scenario. In Section~\ref{sec:matches} we discuss and analyse our performance in all of our matches in the contest. Section~\ref{sec:questions} includes a questionnaire created by the contest organisers with short answers about our team. We end the paper with conclusions in Section~\ref{sec:conclusion}.

\section{Software Architecture}
\label{sec:arch}

In this section we describe the two main software tools that we used to develop our solution to the 2019 \acrshort{mapc}. First, we describe the \acrshort{mas} development platform JaCaMo~\cite{Boissier11} and the features that we made use of to develop our agents and interact with the environment. Then, we discuss the Fast Downward planner~\cite{Helmert06} and how we integrated it into our JaCaMo agents so that they could call their own planner to plan an optimal path.

\subsection{\gls{mas:jacamo}}

The JaCaMo\footnote{\url{http://jacamo.sourceforge.net/}}~\cite{Boissier11} development platform allows \gls{mas} to be programmed at several abstraction layers: organisation, agent, and environment. At the top layer the organisation is defined using Moise~\cite{Hubner07} where groups, roles, links, plan coordination, and norms can be specified. These concepts can be utilised in the middle layer by the Jason~\cite{Bordini07} agents that are programmed following the \gls{bdi} model~\cite{rao91a}. The \gls{bdi} model consists of three mental attitudes: \emph{beliefs} represent the agent's knowledge about the world (including itself), \emph{desires} are goal states that the agent wants to bring about, and \emph{intentions} are partial descriptions of actions to achieve some state. The third layer is defined by CArtAgO~\cite{Ricci09} artifacts. These artifacts are used to describe the environment. An artifact can have \emph{observable properties} that represent the perceptions coming from the environment, and \emph{operations} that describe the outcome of actions in the environment.

Most of the fundamental code architecture in JaCaMo, such as the interface between our agents and the server, has been imported from past participation in the contest from one of our team members, specifically the team PUCRS in 2016~\cite{Cardoso18a}, team 	SMART-JaCaMo in 2017~\cite{Cardoso18b}, and team SMART\_JaCaMo in 2018~\cite{Tabajara19}. We used a snapshot of version 0.7 of JaCaMo. The specific library is available in the \texttt{lib} folder of the repository containing our code.

In terms of the Jason agents we mostly developed from scratch, but still retained some of the architecture from previous years. We kept the use of modules, first used and described in~\cite{Cardoso18b}, to better keep track of beliefs pertaining to each strategy and to help in the debug of the system. We also used the reasoning engine to reconsider agent's intentions, first introduced in~\cite{Tabajara19}. This is useful for agents to be able to change the action that they want to perform before sending it to server. Actions are only sent to the server once all agents in our team have selected an action to perform. It may be the case that an agent selects an action, but then receives information from another agent that causes it to reconsider its own action. We made some small changes to this engine to fit our needs in the new scenario.

In CArtAgO, we use the same structure from previous years. A \gls{team:translator} class is used to translate information from agents to the server (literal to action and terms to parameters) and from the server to the agents (perception to literal and parameters to terms). There is one environment artifact (\gls{team:agentart}) per agent that acts as the agent interface with the server. This artifact also contains the interface with the planner (discussed in the next section). The interface with the server is responsible for registering the agents to the server, collecting and filtering perceptions from the server, and transmitting the actions chosen by the agents. For more details on the functionality of these features we refer the reader to previous papers~\cite{Cardoso18a,Cardoso18b,Tabajara19}.

Apart from the planner interface and updated perception filter (to match the beliefs of the new scenario), the only new addition to the environment artifact is the position of the agent, which is initialised with the (0,0) coordinates. We opted to keep this information in the agent's artifact instead of the agent's belief base because it was easier to keep it consistent after map merges.

Lastly, we have the usual \gls{team:teamart} that is used to share information among all agents in the team. This artifact is very useful to maintain shared data free of race conditions without spending time implementing semaphores and locks to each data structure that agents want to share with the whole team. The \gls{team:teamart} was even more useful than in previous years due to the necessity of sharing (and merging) the maps and the dynamic environment of the new scenario. We discuss the particular operations and observable properties of this artifact in the relevant strategy section (see Section~\ref{sec:strategy} for the descriptions of our strategies).

We have an ad-hoc implementation of roles using the \texttt{my\_role} belief to keep track of which role the agent is playing and some plans that the agents can use to update this belief. Unfortunately, we did not implement these roles in Moise, due to lack of time we had to prioritise other features. Thus, we were not able to use any of the organisation related features that Moise offers in JaCaMo, such as role hierarchy, groups, automatic plans for adopting/abandoning roles/groups, automatic tracking of beliefs related to roles and the organisation, plan coordination, among others. We hope to make use of Moise in the next iterations of the scenario.

\subsection{Fast Downward}\label{sec:fastdownward}

To improve our agents' movement, we used version 19.06 of the
\acrfull{fd} planning system (\cite{Helmert06} and
\cite{Helmert09}\footnote{\url{http://www.fast-downward.org/}}). A
detailed explanation on how we used \acrshort{fd} to improve our
agents' movements is given in Section~\ref{sec:smartmove}, but the
high-level idea is that we wanted to use an out-of-the-shelf planner
able to produce optimal answers within the 4 seconds time limit that agents are required to send their action. We selected \acrshort{fd} for several reasons. First, it is a
well-established planner which has participated several times in the
International Planning Competition (IPC). Second, its results at the
IPC in 2018\footnote{\url{https://ipc2018.bitbucket.io/}} were
promising as the planner performed well in two of the classical
tracks, the satisficing track and the bounded-cost
track\footnote{\url{https://ipc2018-classical.bitbucket.io/\#results}}. Finally,
one of our team members already had some experience with the planner.

The main drawback in using \acrshort{fd} as the chosen planner was the
lack of support for numerical planning. Specifically, while it is
possible in the planning language to express statement such as ``a
clear action requires at least 30 energy to be performed'', the
\acrshort{fd} planner would ignore such information. However, because \acrshort{fd} supports action costs, we were able to devise a workaround the numerical limitation by defining the clear action as the most expensive action (i.e.,
it requires 3 steps to be performed) and then search for plans with minimal action cost. This workaround can
still result in unfeasible plans, but we refer to
Section~\ref{sec:smartmove} for more details.

Due to testing with the planner and the need for fast prototyping,
integration with \gls{mas:jacamo} was rather
rudimentary. Specifically, we implemented a simple shell script to
become familiar with and to test the planner. The script invokes the
planner with the following command line arguments:
\begin{itemize}
\item -{}-sas-file agent\_name.sas : the planner creates a sas file named
  ``agent\_name'' (i.e., each agent has its own file) which
  contains a translation of the planning problem used as input for the
  planner;
\item -{}-plan-file agent\_name\_output : if a plan is returned by the planner, then it
  is written into a file named ``agent\_name\_output'';
\item -{}-search-time-limit 1s : it limits the search time to one second,
  this was necessary in order to stay under the 4 seconds time limit;
\item -{}-alias seq-opt-lmcut : selection of a predefined search
  strategy for optimisation problems that had the best results in our tests; and
\item  the last two parameters were the domain and problem files
  written in PDDL (Planning Domain Definition Language)~\cite{Mcdermott98}.
\end{itemize}
If a plan is returned within the one second time limit, then the
script parses the output file, reprints the plan in a suitable format,
and removes existing sas and output files.

On the \gls{mas:jacamo} side, once an agent decides to use the planner
it checks in the \gls{team:teamart} whether it is allowed to invoke
the planner. This operation performs a simple check of how many agents are currently calling a planner. This was needed when testing our code in slower computers, but during the contest the maximum value of concurrent agents calling a planner was set to 10 (the maximum number of agents in a team), since the computer we were using could comfortably handle 10 simultaneous planner instances. In case of positive answer, it sends its intended goal position to
the \gls{team:agentart}. The \gls{team:agentart} creates an opportune
problem file based on (1) what is in the agent's vision, and (2) what
is the agent's goal. The domain file was created at design time, since it is static and does not change. The script is called as soon as the problem file
is created, and then the script output is parsed. If no plan is found, then the
scripts simply prints ``NO PLAN''. If either ``NO
PLAN'' is returned, or an error occurred during the process, then an
empty plan is returned to the agent. Details on how agents manage
different planner outcomes are described in
Section~\ref{sec:smartmove}.

\section{Strategies and Behaviours}
\label{sec:strategy}

Strategies are what define the main behaviour of our agents. A strategy contains the details on how to solve a particular problem, such as exploring the map or achieving a task. The strategy that each agent uses is based on the beliefs and goals that it has. This is done individually by each agent at key steps in the simulation (i.e., when trigger events of certain strategies occur) and an agent may play multiple strategies in the same step. For example, when an agent is exploring the map and finds another agent of its team both agents momentarily swap to the agent identification strategy. Then, after the identification is complete they revert back to their previous strategy.

\subsection{Exploration}
\label{sec:exploration}

\begin{figure}
    \centering
    \includegraphics[width=0.3\linewidth]{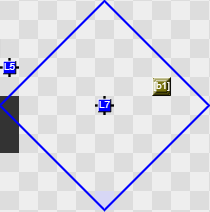}
    \caption{The diamond area represents the vision of agent L7. In this example, L7 perceives that there is an obstacle at (-5,0) and a dispenser for blocks of type b1 at (3,-1). It does not perceive anything else outside this area (e.g., it cannot see agent L5).}
    \label{fig:vision}
\end{figure}

The map used in a round is a randomly generated grid of unknown dimensions. Agents are able to see up to five blocks away, as shown in Figure~\ref{fig:vision}. Any perception received about cells in an agent's vision are oriented using the agent's local coordinates, which is always position (0,0). In other words, at any step in a simulation each agent's local position is (0,0). This remains true even after the agent moves. Therefore, at the start of the simulation it is important to explore the surroundings to try to find goal positions and dispensers of different block types.

There is no use in sharing information about what agents see unless they have access to a global map. However, to be able to create such a map and share information, first these agents have to see each other at least once. Although we have strategies for creating this map (discussed in following sections), for exploration we ignore the global map entirely and each agent only uses its vision to decide what direction to explore (north, south, east or west). Note that using information from the global map (or even the agent's local map) to inform exploration may be more advantageous, but in our tests our current approach proved to be sufficient in exploring most of the map. There is a trade-off between the time necessary for implementing the code for such a complex strategy and a \emph{mostly} random approach to exploration. Although we considered this trade-off not to be worth it, in future contests this should be re-evaluated.

\begin{figure}[ht]
	\centering
	\begin{subfigure}[t]{0.4\textwidth}
        \centering
        \includegraphics[height=1.5in, width=1.8in]{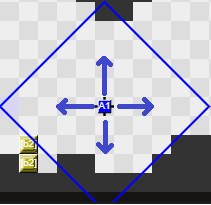}
        \caption{\label{fig:alldirs}Random direction from [north,south,east,west].}
    \end{subfigure}%
    \hspace{0.3cm}
    \begin{subfigure}[t]{0.4\textwidth}
        \centering
        \includegraphics[height=1.5in, width=1.8in]{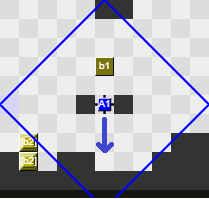}
        \caption{\label{fig:obstructions}Random direction from [south].}
    \end{subfigure}
	\begin{subfigure}[t]{0.4\textwidth}
        \centering
        \includegraphics[height=1.5in, width=1.8in] {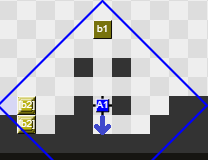}
        \caption{\label{fig:special1}First stage of special exploration.}
    \end{subfigure}%
    \hspace{0.3cm}
    \begin{subfigure}[t]{0.4\textwidth}
        \centering
        \includegraphics[height=1.5in, width=1.8in]{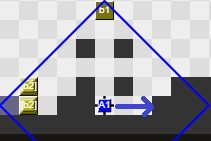}
        \caption{\label{fig:special2}Second stage of special exploration.}
    \end{subfigure}
    \caption{\label{fig:exploration}Different behaviours in exploration.}
\end{figure}

Our agents select a random direction from the list of possible directions, initially containing all four: north, south, east, and west (as exemplified in Figure~\ref{fig:alldirs}). For any subsequent step, the agent will follow the same direction as long as it remains valid. A direction is no longer valid if there is an obstruction at distance less than or equal to 2. For the purposes of this check, we consider an obstruction to be either an obstacle or a block. An example of this is shown in Figure~\ref{fig:obstructions}, where there are two obstacles at (1,0) and (-1,0) of the agent (removing east and  of the list of valid directions), and a block at (0,-2) of the agent (removing north). Therefore, the only remaining valid direction is south.

When the chosen direction is no longer valid, both the direction and the opposite direction are removed from the list of valid directions. For example, if the agent was moving south and encounters an obstruction, both south and north are removed from the list.

If the list of valid directions is empty, then we relax our obstruction condition and call a special case with two stages. In both stages, the obstruction condition is reduced to a distance equal to 1. In the first stage (Figure~\ref{fig:special1}), the agent picks a new direction from the list of valid directions (now repopulated with the relaxed obstruction condition, but still without the opposite direction of the last movement). The agent then moves in the chosen direction until there is an obstruction in the next cell. In the second stage (Figure~\ref{fig:special2}), the agents picks a second direction, where the list of valid directions contains only the opposite axis of the last movement (e.g. if the agent was moving south, possible directions include east and west, assuming they pass the obstruction condition), and again moves in the chosen direction until there is an obstruction in the next cell. After the special case is completed or if it fails, then the agent reverts back to the normal exploration behaviour.

During the normal exploration behaviour (i.e., not in the special case), if agents encounter an obstruction in the direction that they are moving, they will try to execute the \emph{clear} action to remove it as long as they are above a certain \emph{energy threshold}. We used 240 as the threshold, out of a maximum energy of 300. During the contest, a successful clear action had a cost of 30 energy, thus, our agents were able to perform up to two clears while exploring, but still save a reasonable amount of energy for future clears to be used in more important strategies. Longer exploration phases would allow for more clear actions, since agents gained 1 energy per round, up to the maximum amount of energy.

Even though we are using a simple strategy for exploration, there are some conditionals that can be taken into consideration to improve the overall behaviour of the agents that are exploring. For example, it is possible (although somewhat unlikely) that there are teammates in each of the adjacent four directions. In such a case, the agent should attempt to move by selecting one direction randomly out of all four directions.

It is possible (although rare) that the agent will end up with no valid direction to choose from (e.g., an empty direction list) and with an obstruction in all four directions. In such case, the agent will skip a step and then try to clear blocks before reassessing the direction list. Other corner cases include dealing with an out of bounds movement (the agent only knows that it is trying to move outside of the grid when it receives a specific failure code for the movement action) by simply removing that direction from the list, and dealing with an agent from the same team in the immediate direction of movement by trying to go around the agent by choosing the relative right (the other agent will do the same, and since they are moving towards opposite directions, their relative right will be different to each other). We did not have a plan for dealing with agents from the other team. In such case, our agents would simply keep trying to move until being successful.

We considered other viable options for exploration that differ from random direction or movement based on map information, such as picking a single direction for all agents to move or distributing 1/4 of the directions to each agent. However, from our initial tests, selecting a random direction seems to work best across most maps, which is important due to the map randomness present in each simulation.

\subsection{Agent Identification}
\label{sec:id}

In order to cooperate the agents have to first identify each other. Unfortunately, when the game starts the agents do not have this information. When an agent meets another one in its vision it can only recognise if that agent is a member of its own team.
Because of this, one of the first challenges we had to tackle was agent identification. That is, when an agent meets another one of its team, it has to try to identify exactly which agent it is.

Each step, when an agent receives the perceptions from the server, they are checked if there is an unknown entity in its vision which is part of its team. When this happens, a broadcast message is triggered and the agent (let us call it $A_0$) asks to all the agents in its team (let us call them $A_1$, ..., $A_n$) to communicate what they are currently seeing in their vision at this time step. Each agent receiving this broadcast message replies with a list containing all the \texttt{thing} perceptions it has. 

When $A_0$ receives all the replies, the identification process begins. For each reply, $A_0$ analyses the list of things that were reported. First, it has to understand if the reply is from an agent in its vision. For each replying agent $A_i$, $A_0$ checks if in the list of things there is one entity with relative position (\texttt{X}, \texttt{Y}) and in its beliefs there is an entity with relative position (\texttt{-X}, \texttt{-Y}). If this is true, then it means that it is possible that $A_i$ and $A_0$ are currently seeing each other, since both see an entity of their team in the symmetric position. However, this is not enough because there can be multiple agents on the map which are at the same distance and relative position. In order to differentiate these cases, the identification process requires a second step, where for all the things contained in the list, $A_0$ has to check if these are also in its perceptions. Naturally, this is limited to only the things which are in $A_0$'s vision. If after this step only one agent $A_i$ satisfies these constraints, then $A_0$ knows that the agent in its vision is $A_i$. If instead, after this step more than one agent is still suitable for being the agent in vision, then $A_0$ cannot conclude the identity of the entity, and the identification process ends. In the following time steps, until $A_0$ has not identified all the agents in its team, it will keep trying to identify each unknown agent.

\begin{algorithm}[H]
  \textbf{Trigger:} an unknown entity entered my vision\;
  \If{I see an unknown teammate in position (X, Y)}{
  \textbf{broadcast} a request for info to all teammates\;
  \textbf{set} candidates = [ ]\;
  \For{each teammate $A_i$}{
  \textbf{wait} reply containing list L of things seen by $A_i$\;
  \If{thing(-X, -Y, entity, myTeam) \textbf{in} L}{
  possible = \textbf{true}\;
  \For{each thing(W, Z, Type, Name) \textbf{in} L}{
  \If{$\mid$X+W$\mid$ + $\mid$Y+Z$\mid$ $\leq$ 5 \textbf{and} thing(X+W, Y+Z, \_, \_) \textbf{not in} my beliefs}{
  possible = \textbf{false}\;
  \textbf{break}\;
  }
  }
  \If{possible}{
  \textbf{add} $A_i$ \textbf{to} candidates\; 
  }
  }
  }
  \If{length(candidates) == 1}{
  \textbf{add} $A_i$ \textbf{to} the list of identified agents
  }
  }
 \caption{identify()}
 \label{alg:identification}
\end{algorithm}

In Algorithm \ref{alg:identification} we report the pseudo-code for the identification process. The algorithm is triggered when an unknown entity enters the vision of the agent. To simplify the presentation, we assume unification in the conditions, and where we are not interested in the value of a parameter we use the \_ symbol.

\begin{figure}
    \centering
    \includegraphics[width=0.4\linewidth]{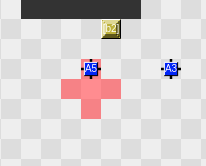}
    \caption{Identification example with two agents and a dispenser.}
    \label{fig:identification}
\end{figure}

In order to better understand how the identification process works, we report a simple example. In this scenario, shown in Figure \ref{fig:identification}, we have two agents of the same team involved, \texttt{A5} and \texttt{A3}. 

For simplicity we focus on the identification process on \texttt{A5}'s side (for \texttt{A3} is symmetric). In this example, we suppose \texttt{A5} has not identified \texttt{A3} yet, and when the belief \texttt{thing(4, 0, entity, ``A'')} is added in its belief base, it broadcasts a request for information to all the agents in its team. When \texttt{A3} receives this request, it sends to \texttt{A5} the list of things is currently seeing in its vision: \texttt{[thing(-4, 0, entity, ``A''), thing(-3, -2, dispenser, b2)]}. With this information, \texttt{A5} first has to check if there is an entity in the list in position \texttt{(X, Y)}, and a corresponding entity in its belief base in position \texttt{(-X, -Y)}. This is satisfied, in fact \texttt{thing(-4, 0, entity, ``A'')} is in the list returned by \texttt{A3}, and in the \texttt{A5}'s belief base, we have \texttt{thing(4, 0, entity, ``A'')}. Thus, it is possible that the agent in \texttt{A5}'s vision is \texttt{A3}. But, to be sure of it, \texttt{A5} also has to check that all the things \texttt{thing(W, Z, Type, Name)}, in the list which are in its vision ($\mid$\texttt{W+X}$\mid$ \texttt{+} $\mid$\texttt{Z+Y}$\mid$ $\leq$ \texttt{5}) are also present in its belief base; meaning that we can find \texttt{thing(W+X, Z+Y, Type, Name)} in the belief base for each thing in the list (where \texttt{X} and \texttt{Y} are the relative coordinates of \texttt{A3} from \texttt{A5}'s viewpoint). 

In this very simple example \texttt{X = 4}, \texttt{Y = 0} and the list contains only \texttt{thing(-3, -2, dispenser, b2)}. Since $\mid$\texttt{-3+4}$\mid$ \texttt{+} $\mid$\texttt{-2+0}$\mid$ = $\mid$ 1 $\mid$ \texttt{+} $\mid$ -2 $\mid$ = 3 $\leq$ \texttt{5}, the dispenser should also be in \texttt{A5}'s vision, and it is, in fact we have \texttt{thing(1, -2, dispenser, b2)} in \texttt{A5}'s belief base. Supposing that there are not other teammates at the same distance, \texttt{A5} can safely conclude that the agent in its vision is \texttt{A3}. The same thing is be done by \texttt{A3}, which will identify \texttt{A5} as the agent in its own vision.

\subsection{Goal Evaluation}
\label{sec:goaleval}

In this section, we present how an agent evaluates a goal position. We need this additional step because we want to pick a suitable goal position to be used when solving the tasks later on. The task completion (see Section \ref{sec:tasks}) is expensive in terms of preparation and communication among the agents; thus, the agents have to ensure that they pick a good goal position. More specifically, a goal position is considered suitable when it has enough space around it for the agents to assemble block structures and to complete the tasks. 

\begin{algorithm}
\KwData{Target position (X2,Y2)}
\textbf{get} agent position (X1,Y1)\;
\While{\textbf{not} (X1,Y1) == (X2,Y2)}{
direction = w\;
distance = $\mid$X1-1-X2$\mid$ + $\mid$Y1-Y2$\mid$\;
\If{$\mid$X1+1-X2$\mid$ + $\mid$Y1-Y2$\mid$ $\leq$ distance}{
direction = e\;
distance =  $\mid$X1+1-X2$\mid$ + $\mid$Y1-Y2$\mid$\;
}
\If{$\mid$X1-X2$\mid$ + $\mid$Y1-1-Y2$\mid$ $\leq$ distance}{
direction = n\;
distance =  $\mid$X1-X2$\mid$ + $\mid$Y1-1-Y2$\mid$\;
}
\If{$\mid$X1-X2$\mid$ + $\mid$Y1+1-Y2$\mid$ $\leq$ distance}{
direction = s\;
distance =  $\mid$X1-X2$\mid$ + $\mid$Y1+1-Y2$\mid$\;
}
\eIf{direction is blocked by obstacle}{
\If{distance $\leq$ 3 \textbf{or} (\textbf{not} go\_around(direction) \textbf{and} enough energy)}{
clear in direction of the target\;
}
\Else{skip\;}
}{move(direction)\;}
\textbf{get} agent position (X1,Y1)\;
}
\caption{move\_to(X2, Y2)}
\label{alg:move_to}
\end{algorithm}

In the goal evaluation process, since the movement required by the agents is not complex, the agents do not call the planner to plan for an optimal path. Calling the planner is computationally expensive, thus it is best to be avoided when possible. For this specific task, we developed a simplified movement algorithm which allows the agents to move and evaluate the goal position.
This algorithm is composed of two parts: direction selection and obstacle avoidance. Assuming the agent has to move to a specific position; in order to choose in which direction the agent should move, it computes the Manhattan distance between its current position and the target destination. Then, the agent moves to the neighbouring cell which minimises the distance that was evaluated. A high-level description of the \texttt{move\_to} function is reported in Algorithm \ref{alg:move_to}. The advantage of this approach lies in being fast, in fact it is not required to compute the entire path in advance, but at each step the agent simply picks the cell which brings it closer to the final destination. Even though this makes the approach very efficient, the presence of obstacles on the path can be a problem. For instance, the closest cell to the final destination can be occupied by an obstacle. 

\begin{algorithm}
\KwData{D is the direction to the obstacle}
\If{D == n \textbf{or} D == s}{
OpposideD = opposite direction of D\;
\If{there is a gap in direction w}{
Ds = [(w,D),(D,e),(e,OppositeD),(OppositeD,w)]\;
}
\ElseIf{there is a gap in direction e}{
Ds = [(e,D),(D,w),(w,OppositeD),(OppositeD,e)]\;
}
\Else{
\Return{\textbf{false}}\;
}
}
\Else{
OpposideD = opposite direction of D\;
\If{there is a gap in direction n}{
Ds = [(n,D),(D,s),(s,OppositeD),(OppositeD,n)]\;
}
\ElseIf{there is a gap in direction s}{
Ds = [(s,D),(D,n),(n,OppositeD),(OppositeD,s)]\;
}
\Else{
\Return{\textbf{false}}\;
}
}

\While{Ds \textbf{is not} empty}{
\textbf{get} and \textbf{remove} first tuple (DtoGo,DTarget) from Ds\;
\While{cell in direction DTarget contains obstacle \textbf{and} new direction to target is not the opposite of DTarget}{
\eIf{cell in direction DtoGo does not contain obstacle}{
move(DtoGo)\;
}{\Return{\textbf{false}}}
}
}
\Return{\textbf{true}}\;
\caption{go\_around(D)}
\label{alg:go_around_obstacle}
\end{algorithm}

We implemented an obstacle avoidance algorithm which allows the agent to go around the obstacles on its path. The high-level description of this is reported in Algorithm \ref{alg:go_around_obstacle}. In this algorithm, the agent knows there is an obstacle in a specific direction $D$, and it wants to go around it. Depending on $D$, the agent creates a projection of the path to avoid the obstacle. Since an obstacle can have two alternative directions to be passed, the agent picks the direction considering the presence of gaps in the path. If the agent sees a gap which could be used to pass the obstacle, it favours that direction over the other. The path projection is coded as a list, which is then used by the agent to move around the obstacle. For brevity and clarity of the presentation, we omit technical details such as: considering action failures and the possibility of having obstacles composed by infinite blocks (although in the contest this never happened).

Now that we have presented the movement used by the agent during goal evaluation, we describe how the actual process of evaluating a goal works. We need to evaluate a goal position due to the complexity of the task completion process. In this evaluation, we check that all the requirements we need to complete the tasks later on are satisfied by the chosen goal position.

The goal evaluation process starts when an agent discovers in its vision a new goal position. First, it has to understand if this goal position belongs to a cluster (set of goals) which has been already evaluated. In order to do that, the agents sharing the same map maintain a list of discovered clusters. A new goal position is added to the closest cluster, i.e., the cluster whose goals are closest to the new one. If no cluster is close enough, it means the new goal belongs to a new cluster and consequently a new cluster containing this single goal position is added to the list. When this happens, it means the agent has discovered a new cluster, which has to be evaluated. The agent that discovered it stops exploring the map and starts evaluating the new cluster.

First, the agent moves to the goal position discovered. Then, it finds the center of the cluster; depending on the form of the cluster there could be multiple centers, in which case one is randomly picked. Once the center is selected, the agent moves to the center. From that position, the agent tries to clear in the 4 cardinal directions, at distance 5 (the maximum in its range). If all the clear attempts succeed, then the goal cluster is suitable since there is enough space around for the task completion (e.g., the positions cleared are not outside the grid of the map). After that, the agent has to find the positions around the cluster which will be used by the retriever agents (the agents fetching the blocks and delivering them to the origin agent later on, see Section~\ref{sec:tasks}). These positions are generated on a rectangle around the goal cluster, and the agent goes in each one of them in order to make sure that they are reachable positions. When the agent finds enough positions for the number of retrievers, in this case 9 (10 agents = 1 origin + 9 retrievers), it stops evaluating the cluster and saves the results.

Even though the map can change because of clear events and actions from other agents, thanks to this evaluation process, we have have a higher probability that a cluster has enough space around it to allow for the completion of complex tasks. The list of retriever positions around the cluster also guarantees that our retrievers will have enough space to move without risking being stuck or trying to move outside the map.

\subsection{Building a Map}
\label{sec:map}

Information about the grid, such as dispenser, goal, and block positions are only received when inside the vision of the agent. Thus, our agents have to save this information in order to keep track of the things that it has discovered. In what follows we describe how agents build their local map, and ultimately a global map that consists of the merging of maps from agents that meet each other while they are moving through the grid.

\subsubsection{Local Map}

Agents can change their position each simulation step; blocks can be moved when attached to an agent or cleared after a clear action/event or after a pattern of blocks is successfully submitted for a task; and obstacles can be cleared. Thus, due to the dynamic environment of the scenario, we only keep the positions of dispensers and goals in the map. Each agents starts with its own local map, with the origin position of the map set to (0,0). Although this local map is saved to the \gls{team:teamart}, initially each agent can only access its own map. There is no reason to access maps from other agents because their coordinates would not match, rendering the information within useless. Overtime, after an agent meets another one for the first time, they will attempt to merge their maps, eventually obtaining a single global map once enough agents have met each other.

Agents also keep track of their global position in their map. This is with respect to the initial origin of the map (0,0). That is, after each step where a successful movement action was performed by an agent, it updates its global position in that map accordingly. For example, at step 0 $A_1$ performs move south; at step 0, $A_1$ perceives that its last action was successful; then, it updates its global position from (0,0) to (0,1).

\subsubsection{Merging Maps}

Two maps are merged when an agent from each map meet. Each agent starts with its own map, we refer to these agents as the leaders of their respective maps. The coordination of the merge happens between the two map leaders. The map that will be merged into another one is decided based on a heuristic: from a list of all the agents, the leader that will coordinate the merge (i.e., the map that will remain) is the leader that appears first in the list (out of these two leaders). For example, consider that we have two agents meet each other $A_1$ and $A_2$, from different maps (respectively $M_1$ and $M_2$). The map $M_1$ has agents [$A_1, A_5$] and $M_2$ has agents [$A_2, A_3, A_4$], where the head of the list is the leader of the map (respectively $A_1$ and $A_2$). First, we identify the leader with priority: from a list of agents [$A_5, A_3, A_2, A_1, A_4$] we know that $A_2$ happens first in the list over $A_1$. This means that $M_1$ will be merged into $M_2$, i.e., the information in $M_1$ will be added to $M_2$ and $M_1$ will be discarded. After the merge, all agents from $M_1$ are now added to the map $M_2$.

\begin{figure}
    \centering
    \includegraphics[width=0.8\linewidth]{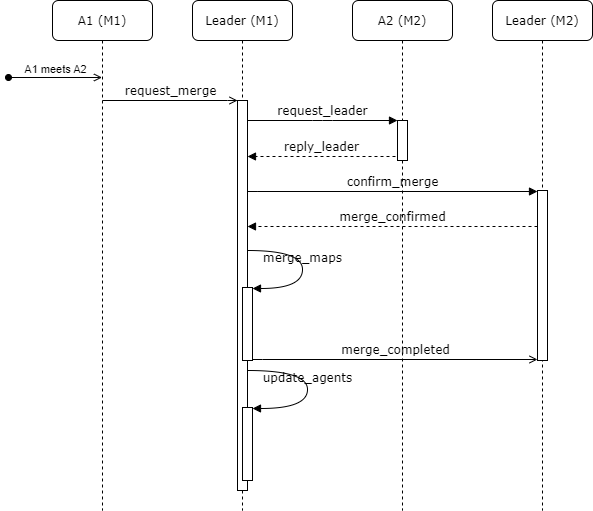}
    \caption{Sequence diagram for the merge protocol. Usual notation from UML sequence diagram applies: solid arrow heads represent synchronous messages, open arrow heads asynchronous messages, dashed lines represent reply messages, and rectangles represent processes.}
    \label{fig:merge}
\end{figure}

We show our merge protocol in Figure~\ref{fig:merge}. It is possible that the leader of map $M1$ is $A1$, in which case the agent sends a message to itself (same result as triggering the intended plan). Similarly, the same applies if the leader of $M2$ is $A2$. It is possible that a merge fails for two reasons: (a) the leader of $M1$ does not have priority over the leader of $M2$ (in which case the merge may be successful when $A2$ tries the merge, instead of $A1$); (b) the leader of $M1$ receives a \texttt{merge\_cancelled} reply instead of a \texttt{merge\_confirmed}, which can be caused by a concurrent merge process which blocked the leader of $M2$, and after this concurrent merge was concluded the leader of $M2$ changed. If a merge fails, agents will proceed to the next simulation step, upon which they may try to merge again (if the merge is still valid).

There are two important stages in the merging of maps, these are shown in Figure~\ref{fig:merge} as the \texttt{merge\_maps} and \texttt{update\_agents} calls. The \texttt{merge\_maps} works as follows:
\begin{enumerate}
    \item \textbf{Calculate new coordinates:} Each dispenser and goal from $M2$ needs to be added to $M1$. However, their coordinates must be updated to consider the origin point of $M1$ instead of $M2$, otherwise their position will not match their real position in the grid. To correctly update these coordinates, we simply have to add to their original $X$ and $Y$ values what we call the new origin coordinates $NewX$ and $NewY$ respectively. We calculate these values as follows: $NewX = (GlobalA1X + LocalA2A1X) - GlobalA2X$, wherein $GlobalA1X$ is the global position of $A1$ in the X axis of $M1$, $LocalA2A1X$ is the position of $A2$ in the X axis according to the local view of $A1$, and $GlobalA2X$ is the global position of $A2$ in the X axis of $M2$. The same is calculated for the $NewY$ value using the Y axis.
    \item \textbf{Update artifact:} Each new position that was calculated is added in the \gls{team:teamart} map structure for $M1$.
\end{enumerate}

The \texttt{update\_agents} works as follows:
\begin{enumerate}
    \item \textbf{Build new list of identified agents:} the new list of identified agents includes all agents that are a part of $M1$ and $M2$. This will be the new list of members of $M1$.
    \item \textbf{Send update to the leader of $M2$:} the new list of identified agents and the new origin coordinates $(NewX, NewY)$ are sent to the leader of $M2$, which uses it to update its global position in the new map. This message is sent separately to the leader of $M2$ purely for bookkeeping purposes.
    \item \textbf{Send update to all agents of $M1$:} the new list of identified agents is sent to the members of $M1$ from before the merge. These agents are staying in their original map, and therefore, they do not need to update their global positions.
    \item \textbf{Send update to all agents of $M2$:} the new list of identified agents and the new origin coordinates $(NewX, NewY)$ are sent to each member of $M2$, which are used to update their global position in the new map.
\end{enumerate}

\subsection{From Exploration to Tasks}
\label{sec:stop}

Naturally, our agents cannot explore the map forever, and eventually they have to start focusing on completing tasks. The decision of stopping exploration and swapping to task completion can be triggered by two different events: the addition of a new task, or, the merging of two maps.
In the first case, the agent has to check it the following conditions are satisfied:
\begin{itemize}
    \item At least one goal cluster has been found and evaluated positively.
    \item The agent has identified at least 2 teammates.
    \item For each block required to complete any one task, the agent knows a dispenser to generate that type of block.
\end{itemize}
If all the previous conditions are satisfied, it means the agent has gathered enough information from its exploration, and it can start solving tasks. 

It is possible that upon the time an agent decides to stop, not all the agents are in the same group (i.e., not all have been identified by the agent which stops). Nonetheless, waiting to have all the agents in the group would be too strict, and would make the approach too slow. Because of this, the group with the agent which has stopped does not wait for the other agents to join it. This brings us to the second event which can make an agent stop exploring, namely when it joins a group which has already stopped. Agents of other groups will keep exploring and eventually will meet an agent of the group that has stopped, and when this happens, they will stop exploring and start retrieving blocks for achieving the tasks.

The first agent which stops exploring becomes the origin for the task resolution. Since the addition of a task is perceived by all the agents, it is possible that multiple agents decide to stop at the same time, but only one will be elected to be origin, while the others will become retrievers. The retrievers and the task origin are discussed further in Section~\ref{sec:tasks}.

\subsection{Optimal In-Vision Path Planning}
\label{sec:smartmove}
This section describes how agents move in the map once the exploration
phase is over. The difference in behaviour between moving during
exploration and moving afterwards is due to different focus on the
agents' side. Specifically, exploration agents are only concerned in
gathering enough information to start performing their tasks, but
after the exploration each agent has a specific objective (e.g.,
retrieving a particular block, or going to a goal position). This
implies that after the exploration phase agents have clear
destinations to reach. As the map changes constantly due to factors
such as clear events or enemy agents moving blocks, we decided to
avoid an approach based on what the agents believe the map to look
like, and to pursue an approach based on the local vision of
agents. In what follows we present in detail the scenario where an
agent wants to reach a specific destination (e.g., next to a dispenser
or next to a goal position). The scenario is divided into three parts:
selection of a destination; invocation of the \acrshort{fd} planner;
and managing planner result.

\subsubsection{Selection of a destination}
Based on the agent's task, the agent decides to reach a specific
destination (given in global coordinates). The agent translates the
global coordinates into coordinates relative to its current
position. Most likely, the resulting coordinates are not within the
agent's vision, and to reach such a destination the agent recursively
plans a path to temporary destinations, bringing it closer to the
final destination. The selection of temporary destinations is based
on heuristics, which try to select `good cells'. In what follows we
refer to a cell as a \emph{good cell} if there is neither an agent nor
a block at the time of evaluation. The heuristic used to select a
temporary destination is as follows. First, the agent computes the
relative coordinates $(x, y)$ of a cell at the border of its vision
(i.e., $\lvert x \rvert + \lvert y \vert = 5$), which represents the
cell closest to the destination. If the computed cell is a good cell,
then the agent selects such cell as its temporary
destination. Otherwise, the agent tries to move either north (resp.,
south) or east (resp., west) based on whether
$\lvert x \rvert < \lvert y \rvert$ and $y < 0$ (resp., $y > 0$), or
$\lvert x \rvert > \lvert y \rvert$ and $x > 0$ (resp., $x < 0$). In
case it decides to move north (resp., south), it evaluates the cell
$(0, -5)$ (resp., $(0, 5)$). The behaviour for moving east (resp.,
west) is analogous. If the cell is a good cell, then it becomes the
temporary destination. Otherwise, the process is repeated for the
three closest cells. If no good cell is found during this process,
then the agent behaves in the same way as when it receives an empty
plan, which is described later on in this section.

\subsubsection{Invocation of the \acrshort{fd} planner}
In the eventuality that a destination has been selected by the agent,
then it invokes the planner in order to plan its next few
actions. When invoking the planner, several details need to be
provided by the invoking agent. First, the agent has to specify
whether or not a block is attached, and what its relative coordinates
are. Second, the agent has to specify what is its goal (e.g., the
agent wants to reach the destination, or the attached block has to
reach the destination). Finally, whether or not the planner is allowed
to return a plan using the clear action. The latter information is due
to the \acrshort{fd} planner not handling numerical planning, but
handling action cost optimisation. As mentioned in
Section~\ref{sec:fastdownward}, the heuristic we used is based on the
idea that the clear action is the most expensive action since it is
the only one requiring three steps. Hence, we regarded unlikely for
the planner to return an optimal solution containing more than one
clear action. For this reason the agent only checks whether or not it
has enough energy to perform at least one clear action. In case of a
positive answer the planner is invoked with the possibility of using
clear actions, and otherwise such a possibility is
forbidden\footnote{From a modelling point of view, allowing or not a
  clear action is based simply on using a PDDL domain file with or
  without the definition of the action.}. All the aforementioned
information is passed to the \gls{team:agentart}. The
\gls{team:agentart} already contains information on what is in the
agent's vision, and uses such information and the details provided by
the agent in order to create a new PDDL problem file. Since we did not
want to perform any clear action on blocks attached to allied agents,
the domain files do not allow the planner to perform clear actions
over blocks. We also did not want to perform clear actions on agents,
and decided to treat them as blocks. As soon as the
\gls{team:agentart} has created the PDDL problem file, then it invokes
the shell script with the call to the planner and parses its
output. If an error occurred during the process of problem creation,
or if the planner did not return a solution (i.e., no solution exists
or no plan was found within the one second timeout), then an empty
plan is returned to the agent. Otherwise, a sequence of actions is
returned.

\subsubsection{Managing planner result}
Any returned solution (empty or not) is processed by the agent. If a
sequence of actions is returned, then the agent executes them in a
rather blind fashion. It can, of course, happen that some action fails
(e.g., a random fail occurs, or an unexpected obstacle appears). If
the failing action is not a movement action, then the event is simply
ignored. Otherwise, the agent tracks the failed movement in order to
have an up-to-date information regarding its distance to the
destination. There are two reasons for such a forgiving approach to
action failures. First, re-invoking the planner can easily result in a
huge consumption of resources because the returned plan is based on a
snapshot of the agent's vision, and the dynamic nature of the
environment will easily lead to new faulty plans. Second, the agent is
getting closer to its destination as long as not all planned actions
fail. 
The situation is different when the returned solution is empty (or if
the destination selection heuristic does not find a good cell) since
the agent cannot delegate its decision making process to the planner,
but it is still required to take a decision. In this case the agent
tries to perform a single step movement action before invoking the
planner again. The direction of the movement is again based on
heuristics involving some of the same ideas as those used for
destination selection. The reasoning behind performing one movement
action before a new planner invocation is the low likelihood of the
situation to stay exactly the same. In other words, we thought that a
single action together with the dynamic nature of the environment was
enough of a change, which should result in increasing the likelihood
for a new planner call to be successful. The heuristics we used
resulted in a good performance of our system in this year competition,
but they are clearly not optimal yet. In fact, there is at least one
example where our heuristics resulted in a loop in an agent's
behaviour. Specifically, the loop described in
Section~\ref{sec:matches} in round two against the team~\gls{dtu}
is the result of the environment not changing enough after one time step.

\subsection{Achieving Tasks}
\label{sec:tasks}

We only consider tasks in the time step that they have been announced by the server. From experience in past contests, in particular~\cite{Cardoso18b,Tabajara19}, we know that it is more advantageous to first build a stock of items that we may require to complete tasks. Then, once we have everything we need to complete the task, we commit to trying to achieve it. A team that uses this strategy of stocking items will always complete a task before a team that first decides on a task and then tries to obtain the necessary items. Since teams share the same pool of tasks, failing a task because the other team completed it first can be very detrimental to the score of the team, since it is very hard to reuse the resources of the failed task to a new task.

Agents that decide to stop exploring will switch their strategy to pursue and achieve tasks. The first agent in the group that decides to stop becomes the \emph{task origin}. The task origin is responsible for:
\begin{enumerate}
    \item choose a good cluster goal;
    \item move to the appropriate goal position within the cluster;
    \item clear any blocks or obstacles while idle (i.e., not trying to achieve a particular task);
    \item evaluate and select a new task to be achieved;
    \item coordinate with the other agents to build the pattern required by the task;
    \item submit the task.
\end{enumerate}

The remaining agents from the group will become the retrievers. Retrievers are responsible for building a stock of blocks that can later be used to build block patterns. Retrievers will pick a block from a pool of block types, which is constantly updated to ensure that we have a good variety of blocks available in our stock. After collecting a block, the retriever will move to a designated position (previously scouted during goal evaluation) where it waits for a block request from the task origin.

A task will be selected only if it has been announced in the active step and if there are enough retrievers in position with all of the blocks that the task requires. The task origin contacts the appropriate retrievers to request them to bring their blocks. While some retrievers may bring their blocks in parallel, this is only allowed when it is not possible to have a conflict. In other words, multiple retrievers can bring their blocks in parallel only if the designated position in the pattern does not require a block that another retriever is currently bringing. This constraint is only necessary to alleviate some of the burden from coordinating the agents, and could be removed if proper coordination was in place. Once a retriever adds his block to the pattern, it reverts to its normal behaviour and goes to fetch another block for the stock. After the pattern is complete, the task origin submits the task.

The clear downside of having only one task origin is that it is a single point of failure. It worked in our favour because this was the first contest with a new scenario and the other teams didn't try to interfere too much with agents from the opposing team.


\section{Match Analysis}
\label{sec:matches}

In this section we analyse all of our matches in the 2019 \acrshort{mapc}. There were a total of four participating teams: \gls{lfc}, \gls{fit}, \gls{trg}, and \gls{dtu}. Each team used a different programming language (respectively): JaCaMo, Java, Jason, and GOAL. In total, our team won seven simulation rounds, had one draw, and lost one. With these results, we achieved a total of 22 points (3 per win, 1 per drawn, 0 per loss) and got the first place in the contest (7 points ahead of the second team, \gls{fit}).

In Table~\ref{tab:results}, we show the total score obtained by each team. The total score is the sum of the score obtained from successfully delivering tasks across all rounds. The highest score in a single round was achieved by team \gls{fit}, with a score of 680. In that particular round, their agents managed to deliver several tasks of size three (i.e., three attached blocks). Although \gls{trg} generated more overall score than \gls{dtu}, their final placement is lower (fourth place) since they only won one round in contrast with the three wins from \gls{dtu}.

\begin{table}[ht]
	\centering
	\caption{Total score of each team.}
	\begin{tabular} {l@{\hspace{2em}} c@{\hspace{2em}}} \toprule
		Team & Total Score\\ \midrule
		\gls{lfc}               & 	1790  \\
		\gls{fit}             & 1760 \\ 
		\gls{trg}              & 590 \\
		\gls{dtu}              & 330  \\ 
		\bottomrule
	\end{tabular}    
	\label{tab:results}
\end{table}

Due to the inherent randomness of this year's scenario, we are not able to do direct comparisons between rounds 1, 2, and 3, in each of the matches since the map in a round can be completely different in the same round of another match. 

\subsection{\gls{lfc} vs \gls{fit}} 

Our first match was against \gls{fit} and it was decisive in determining the winner of the contest. Although \gls{fit} performed very well against the other teams, achieving a total score in delivered tasks similar to ours, they were not able to complete any tasks (except for one in the third round) against our team. Our agents did not (intentionally) try to sabotage the opponents in any of our matches. After a task is successfully delivered or if it fails, our origin agent tries to clear any visible cell within its vision radius of any blocks/obstacles. All agents will also perform a similar clear if they reconnect to a match (due to latency or deadlocks). We cleared some of the blocks from \gls{fit} agents in this way, however, it is difficult to ascertain if that is what influenced our victories in any of the three rounds.

During the first match, we noticed a bug that would occur after successfully delivering a task where the origin agent would (sometimes) send no action for several steps. Occasionally the agent was able to recover from it by itself after a few steps, otherwise we had to manually reconnect all agents. This had a severe impact on the amount of tasks we could complete in the round, since after a reconnect our team started from scratch. We believed the source of this bug was the strategy for failure recovery that we added in the last few days before the contest. After the match, we decided to disable this feature for future matches. This did not have the desired impact, as the bug still occurred and our team was not able to automatically recover from failure in the remaining matches.

\subsection{\gls{lfc} vs \gls{trg}}

We won the first and the third round and had a draw in the second round against \gls{trg}. \gls{trg} observable tactics consisted in splitting the agents into two groups. One focused on solving the tasks, and the other one (majority) focused on patrolling goal clusters (possibly to stop the other team from completing tasks). The agents belonging to the latter occupied the goal positions not used by they teammates, and they kept patrolling these areas clearing all the blocks in their vision. 

In the first round, our team quickly achieved a score of 180 by step 256 (out of 500), which was our final score in that round. Due to pre-existing blocks close to the origin position, one of our agents that was bringing a block to the origin got stuck in a movement loop. Since we removed failure recovery after the previous match against \gls{fit}, our agents never recovered. Although we were watching the matches live, we could only see the results of each agent's action, and thus did not notice the bug. We could have easily lost/tied the round due to this bug, since \gls{trg} performed very well and their agents were still completing tasks after our agents stopped. The third round was very similar to the first, except this time we noticed the bug and tried to reconnect the agents. Unfortunately, they were not able to deliver any tasks after reconnecting, however, we already had a good lead over \gls{trg}, enough to secure our win in that round.

In the second round, the patrolling \gls{trg} agents picked the goal cluster that our agents selected to deliver tasks. Thus, each time our agents tried to complete a task, their agents cleared the blocks attached to our origin agent, causing our agents to fail. Since we disabled automatic failure recovery, we were unable to deliver any tasks in this round. Luckily, \gls{trg} also did not manage to deliver any tasks, thus the round ended in a draw. This round was a good example of our single point of failure, the origin agent. Since we only use one goal cluster, if this cluster is disputed in any way, our team fails and is not able to properly recover from it. Future extensions should consider using multiple origin agents or having dynamic movement between different clusters.

\subsection{\gls{lfc} vs \gls{dtu}}

\gls{dtu} is the only team to which we have lost a round. Specifically, in the
second round \gls{dtu} won by 130 to 40, while our team won rounds one and three by a good margin. By observing the
match against \gls{dtu}, we noticed the following behaviour. First, their agents explore the map to collect as many blocks
as possible (i.e., each agent collects four blocks). Second, as
soon as the agents agree on a task, then those involved in the task
detach from unnecessary blocks, and move with the only remaining block
into a goal position. Finally, the agents try to assemble and to
submit the task. As any strategy, the one adopted by \gls{dtu} has
advantages and disadvantages. On one side, their strategy seems to be
rather resilient to potential attacks by other teams since the goal
position is established only at the time of completing a task and each agent carries multiple blocks. On the
other hand, having all the agents moving around with four blocks
attached to them is not trivial. For example, in the replay of
round one around step 305 agent \emph{G5} collects four blocks, but then it
is unable to move away from the dispenser because the attached blocks
collide with surrounding obstacles and it is stuck on a loop for the remaining of the round. This is why we decided that our agents would only move with one attached block.

Round one and three had the same score and played very similar. We won both of them
with a resulting score of 380 to 40. Our performance in both rounds were a result of the limited
interaction between agents of different teams (i.e., for the most part there was no conflict in goal clusters), and a faster task
execution by \gls{lfc} after the exploration phase. 

Round two,
however, is more interesting to analyse because weaknesses of our approach were clearly shown. In such round,
our team scored 40 points and we had to reconnect the
agents twice. A reconnect means that all knowledge
acquired by the agents is lost, and all agents resume the initial strategy of exploring the map after clearing the surrounding area. The
first reset happens at step 190, when we realised that agent \emph{L6} was
stuck in a loop. The loop was caused by the enemy agent \emph{G9} being in
the cell which \emph{L6} consider its destination. This resulted in agent \emph{L6}
moving back and forth hoping for a change in the situation (e.g., that \emph{G9} would move to a different cell). As agent
\emph{L6} was crucial for the completion of the task, and the situation was
not changing, we opted to reconnect our agents. After the first
reconnect, it took almost 100 steps for the agents to gather enough
information and stop their exploration. The agents behaved as expected
until a clear event at step 304 occurred. The event resulted in our
origin to be disabled, making any attempt of fulfilling a task
unsuccessful. It took us about 100 steps before recognising the
problem, and the second restart of the agents occurred at step
408, which was clearly too late for our agents to recover from the score
disadvantage, resulting in our only loss in the contest. 

\section{Team Overview: Short Answers}
\label{sec:questions}

\subsection{Participants and their background}
\begin{description}
\item \vskip0.5em\textbf{What was your motivation to participate in the contest?}
\\
The new scenario introduced this year was our main motivation, since it required a fresh start and a new perspective to solve it due to the additional randomness of its grid environment. We also wanted to improve our knowledge about agent technologies, and put them into practice in a complex multi-agent scenario.
\item \vskip0.5em\textbf{What is the history of your group? (course project, thesis, $\ldots$)}
\\
All members of our group are post-doctoral research associates at the University of Liverpool.
\item \vskip0.5em\textbf{What is your field of research? Which work therein is related?}
\\
Our work at the moment mostly relates to formal verification. We have not formally verified any aspect of our implementation yet, however there are several things that could be interesting to verify, such as our protocol to merge the maps, the individual planning of the agents, etc.
\end{description}

\subsection{Statistics}
\begin{description}
\item \vskip0.5em\textbf{How much time did you invest in the contest (for programming, organizing your group, other)?}
\\
Approximately 200 hours.
\item \vskip0.5em\textbf{How many lines of code did you produce for your final agent team?}
\\
A total of 6,783 lines of code, with 1,369 in Java and 5,414 in Jason.
\item \vskip0.5em\textbf{How many people were involved?}
\\
Three post-docs:
\\ Rafael C. Cardoso (PhD, Postdoctoral Research Associate at University of Liverpool)
\\ Angelo Ferrando (PhD, Postdoctoral Research Associate at University of Liverpool)
\\ Fabio Papacchini (PhD, Postdoctoral Research Associate at University of Liverpool)
\item \vskip0.5em\textbf{When did you start working on your agents?}
\\
Our first commit to our online repository was on May 8th, 2019. However, most of the work started on September and continued up to the day of the contest.
\end{description}

\subsection{Agent system details}
\begin{description}
\item \vskip0.5em\textbf{How does the team work together? (i.e. coordination, information sharing, ...) How decentralised is your approach?}
\\
Our agents start completely decentralised, each with their own local map. As the agents meet each other, they start merging their information. When evaluating tasks and assembling a team, our solution is centralised, as only the agent in the origin goal position can do so.
\item \vskip0.5em\textbf{Do your agents make use of the following features: Planning, Learning, Organisations, Norms? If so, please elaborate briefly.}
\\
Each agent uses an instance of an automated planner (Fast Downward) to plan its movement using only its local vision. If the target position is outside of its local vision, the agent picks the closest position inside its vision and plan a path to go there, once it arrives it loops until arriving at the desired target position.
\item \vskip0.5em\textbf{Can your agents change their behavior during runtime? If so, what triggers the changes?}
\\
Initially all agents are explorers, but when enough agents meet each other and enough dispensers and goal positions are known, then agents start to migrate to different roles, such as goal origin, block retriever, etc.
\item \vskip0.5em\textbf{Did you have to make changes to the team (e.g. fix critical bugs) during the contest?}
\\
Close to the start of the first match we noticed that we had a bug where after completing a task the origin agent would die (i.e. send no action) for a few steps. We couldn't change anything in time for the first match, but we tried fixing it while the match was underway. Unfortunately, our changes ended up making our team worse and not fixing the bug.
\item \vskip0.5em\textbf{How did you go about debugging your system?}
\\
Unfortunately JaCaMo does not have good debugging features yet, thus we used prints to test that critical parts of our code were behaving as intended.
\item \vskip0.5em\textbf{During the contest you were not allowed to watch the matches. How did you understand what your team of agents was doing? Did this understanding help you to improve your team's performance?}
\\
We were only able to see when one of our agents died (i.e. sent no action), but without the map it was difficult to understand the context of the problem. For example, during one of the matches we noticed (via prints) that one of our agents was stuck in a loop, however, we were not able to tell if this agent was involved in performing a task or not. Without this knowledge, we decided to reconnect all agents, which is very costly as they start from zero.
\item \vskip0.5em\textbf{Did you invest time in making your agents more robust? How?}
\\
Only very close to the day of the contest did we add task failure recovery and automated round change, until then we were improving our main strategies. This is probably what caused most bugs, since we did not have time to test it properly.
\end{description}

\subsection{Scenario and Strategy}
\begin{description}
\item \vskip0.5em\textbf{What is the main strategy of your agent team?}
\\ 
Our main strategy is twofold: first we build a consistent map that can be used by all agents to speed up the achievement of new tasks; then we move smartly using an automated planner.
\item \vskip0.5em\textbf{Your agents only got local perceptions of the whole scenario. Did your agents try to build a global view of the scenario for a specific purpose? If so, describe it briefly.}
\\
Our agents merge their local maps when they meet a new agent. This map contains limited information: the coordinates of dispensers and goal positions. Each agent maintains its own position in the current map up to date, but this is not shared unless specifically requested by another agent.
\item \vskip0.5em\textbf{How do your agents decide which tasks to complete?}
\\
An agent moves to a scouted origin position (must be a goal position). Once in position, this agent evaluates any new task to check if there are enough agents in place (nearby the origin) with the correct block types for the task.
\item \vskip0.5em\textbf{Do your agents form ad-hoc teams to complete a task?}
\\
The team is formed by the origin agent, the other agents simply receive a subtask to bring their block to a particular position.
\item \vskip0.5em\textbf{Which aspect(s) of the scenario did you find particularly challenging?}
\\
A major barrier at the start was the lack of shared information between the agents. Since each agent had its own local vision, we had to come up with a solution so that agents could identify each other when they met, and then exchange information to start building a shared map. Another challenge was to deal with the dynamic clear events coming from the environment, as these could create/remove obstacles and erase blocks that our agents were using to complete a task.
\item \vskip0.5em\textbf{If another developer needs to integrate your techniques into their code (i.e., same programming language tools), how easy is it to make that integration work?}
\\
If the same programming language is used (JaCaMo) then it should be easy to add some of our techniques, due to the modularity of our solution.
\end{description}

\subsection{And the moral of it is \ldots}
\begin{description}
\item \vskip0.5em\textbf{What did you learn from participating in the contest?}
\\
Combining classical automated planning with agents is not as slow as we thought it would be. Of course the problem had to be reduced to include only the local vision of the agent, but given the timeout of 4 seconds, we were surprised that all 10 agents were able to plan concurrently in the same step.
\item \vskip0.5em\textbf{What are the strong and weak points of your team?}
\\
The strong points of our team are the map merging techniques and the individual path planning. Our weak point is that we only have one agent in an origin goal position, so if something goes wrong with it we are unable to complete tasks.
\item \vskip0.5em\textbf{Where did you benefit from your chosen programming language, methodology, tools, and algorithms?}
\\
We used the JaCaMo development platform, a combination of Jason, CArtAgO, and Moise. We benefited from the modularity and the many internal actions available in Jason. CArtAgO helped us interface with the server and share common knowledge between our agents. We did not use Moise, but we hope to do so next year.
\item \vskip0.5em\textbf{Which problems did you encounter because of your chosen technologies?}
\\
Debugging a platform that uses multiple technologies is complicated. Most of our tests had to be done by adding prints in several places we wanted to test.
\item \vskip0.5em\textbf{Did you encounter new problems during the contest?}
\\
We encountered several problems. The most troublesome was that after delivering a task the origin agent would die (i.e. send no action) for a few steps. While trying to fix this problem we deactivated things such as failure recovery from the second match forward. This made our team worse and did not fix the problem.
\item \vskip0.5em\textbf{Did playing against other agent teams bring about new insights on your own agents?}
\\
There were many interesting strategies used by other teams. TRG had a team of disruptive agents that would keep moving in a goal cluster. GOAL-DTU made use of all four sides of an agent to connect four blocks. FIT BUT agents would meet with complex block structures and assemble it close to a goal cluster, then after the structure was complete they would move in a goal position to deliver.
\item \vskip0.5em\textbf{What would you improve (wrt. your agents) if you wanted to participate in the same contest a week from now (or next year)?}
\\
We would make use of Moise, since most problems we had were related to coordination between agents, something that Moise organisations excels at.
\item \vskip0.5em\textbf{Which aspect of your team cost you the most time?}
\\
We had long discussions about how to merge maps effectively, and the planning component took some time to integrate with the agents.
\item \vskip0.5em\textbf{What can be improved regarding the contest/scenario for next year?}
\\
I think longer matches would be better. While 500 steps make it faster to play, I think they were too short to really display the best characteristics of each team. Having more agents per team will also enable more interesting strategies. Finally, it would be interesting to have actions that could interact/interfere with agents of the opposing team.
\item \vskip0.5em\textbf{Why did your team perform as it did? Why did the other teams perform better/worse than you did?}
\\
We believe our success was due to the combination of: our agent identification, map merging, task coordination, and movement using an automated planner. Due to our centralised solution in solving tasks, our team was very vulnerable to any attempt that would disrupt our agent in the goal origin. Luckily, the other teams were also focused on solving the tasks.
\end{description}
\section{Conclusion}
\label{sec:conclusion}

In this paper we have described the tools and strategies used by our team (\acrshort{lfc}) to win the 2019 \acrshort{mapc}. Our main contribution is the integration of JaCaMo agents with the Fast Downward planner. Although the four second deadline per step can often restrict the techniques used, we were able to workaround this limitation by reducing the state space that the planner had to search, while still allowing each agent to call its own instance of the planner without surpassing the deadline. Our strategies for constructing a global map (including exploration and agent identification) and achieving tasks (including choosing a goal position, retrieving blocks, and assembling block structures) were also crucial in achieving our results. 

Since the number of agents per team will likely increase for the next contest, we would like to improve the interaction between agent and planner to be less computationally demanding. Another important improvement would be to revamp the model of the problem that is translated from the agent's knowledge and given as input to the planner to include more details about the environment. Future iterations of our team should also make use of Moise to program a role-based organisation and to implement proper coordination among agents.

\bibliographystyle{splncs04} 
\bibliography{lfc}  

\end{document}